\documentclass[
 amsmath,amssymb,
 aps,
prb,twocolumn, superscriptaddress]{revtex4-2}

\usepackage{graphicx}
\usepackage{dcolumn}
\usepackage{bm}
\usepackage{xcolor}
\usepackage[colorlinks= true,urlcolor=blue,linkcolor= blue,citecolor=blue,bookmarks=false,pdfstartview=]{hyperref}

\newcommand{\cuchch}{[Cu(pym)(H$_2$O)$_4$]SiF$_6$·H$_2$O}

\begin{document}


\title{Magnetic-field-induced ordering in a spin-1/2 chiral chain} 

\author{Rebecca Scatena}
 \affiliation{Diamond Light Source, Harwell Science and Innovation Campus, Didcot, OX11 0DE, United Kingdom}

\author{Alberto Hern\'{a}ndez-Meli\'{a}n}
 \affiliation{Department of Physics, Durham University, South Road, Durham, DH1 3LE, United Kingdom}

 \author{Benjamin M. Huddart}
 \affiliation{Department of Physics, Durham University, South Road, Durham, DH1 3LE, United Kingdom}
\affiliation{Clarendon Laboratory, Department of Physics, University of Oxford, Parks Road, Oxford, OX1 3PU, United Kingdom}

 \author{Sam Curley}
 \affiliation{Department of Physics, University of Warwick, Coventry CV4 7AL, United Kingdom}

 \author{Robert Williams}
 \affiliation{Department of Physics, University of Warwick, Coventry CV4 7AL, United Kingdom}

\author{Pascal Manuel}
\affiliation{ISIS Pulsed Neutron Source, STFC Rutherford Appleton Laboratory,
Didcot, Oxfordshire OX11 0QX, United Kingdom}

\author{Stephen J. Blundell}
\affiliation{Clarendon Laboratory, Department of Physics, University of Oxford, Parks Road, Oxford, OX1 3PU, United Kingdom} 

\author{Zurab Guguchia}
\affiliation{PSI Center for Neutron and Muon Sciences (CNM), 5232 Villigen PSI, Switzerland}

 \author{Zachary E. Manson}
\affiliation{Department of Chemistry and Biochemistry, Eastern Washington University, Cheney, Washington 99004, USA}
 
\author{Jamie L. Manson}
\thanks{Deceased 7 June 2023.}
 \affiliation{Department of Chemistry and Biochemistry, Eastern Washington University, Cheney, Washington 99004, USA}

\author{G. Timothy Noe}  
\affiliation{National High Magnetic Field Laboratory (NHMFL),
Los Alamos National Laboratory, Los Alamos, NM, USA}
 
\author{John Singleton}  
\affiliation{National High Magnetic Field Laboratory (NHMFL),
Los Alamos National Laboratory, Los Alamos, NM, USA}

 \author{Tom Lancaster}
 \affiliation{Department of Physics, Durham University, South Road, Durham, DH1 3LE, United Kingdom}

 \author{Paul A. Goddard}
 \email{p.goddard@warwick.ac.uk}
 \affiliation{Department of Physics, University of Warwick, Gibbet Hill Road, Coventry, CV4 7AL, United Kingdom}

\author{Roger D. Johnson}
 \affiliation{Department of Physics and Astronomy, University College London, Gower Street, London, WC1E 6BT United Kingdom}
\affiliation{London Centre for Nanotechnology, University College London, Gordon Street, London WC1H 0AH, United Kingdom}

\date{\today}

\begin{abstract}
We present neutron diffraction, muon spin rotation and pulsed-field magnetometry measurements on the Heisenberg quantum chiral chain \cuchch, which displays a four-fold-periodic rotation of the local environment around the Cu(II) $S = 1/2$ ions from site to site along the chain. Previous measurements on this material have shown the absence of magnetic order down to surprisingly low temperatures $\ge 20$\,mK, as well as the presence of an energy gap for magnetic excitations that grows linearly with magnetic field. Here we find evidence at dilution refrigerator temperatures for a field-induced transition to long-range magnetic order above an applied magnetic field of 3\,T. From the polarization of magnetic moments observed in applied fields we can identify the static magnetic structure that best accounts for the data. The proposed model is supported microscopically by the presence of an alternating component of the $g$ tensor, which produces an internal two-fold staggered field that dictates both the direction of the ordered moments and the effective coupling between adjacent chains. The observed magnetic structure is contrary to previous proposals for the departure of the magnitude and field dependence of the energy gap from the predictions of the sine-Gordon model.
 \end{abstract}

\maketitle

\section{Introduction}
The remarkable effects of an alternating local spin environment on the magnetic properties of quantum spin chains were first noticed in high-field neutron scattering and heat capacity data taken on the quasi-one-dimensional staggered $S = 1/2$ antiferromagnetic (AFM) chain, copper-benzoate (Cu(C$_6$H$_5$COO)$_2\cdot$3H$_2$O)
~\cite{Dender1997}. These revealed the development of an energy gap which grows on application of magnetic field ($H$). This was perplexing until, in a theoretical {\it tour de force}, it was shown that the behavior of this system could be described by the sine-Gordon (SG) model of quantum field theory~\cite{Oshikawa1997a,Affleck1999}. Over the intervening years, several other SG chains emerged, including Cu-PM ([pym·Cu(NO$_3$)$_2$·(H$_2$O)$_2$]$_n$, pym = pyrimidine) and CDC (CuCl$_2\cdot$2((CH$_3$)$_2$SO))~\cite{Feyerherm2000,Helfrich1998,Chen2007}. Under the influence of an applied field, the gap emerges in these systems thanks to the presence of an alternating, or two-fold-periodic staggered field with a component that lies perpendicular to the applied field, which is caused by the combination of a two-fold staggered component in the Cu(II) $g$ tensor and an alternating Dzyaloshinskii-Moriya (DM) interaction~\cite{Oshikawa1997a,Affleck1999}. Soliton, anti-soliton and breather excitation modes were predicted~\cite{Affleck1999,Essler1998,Essler2003} and then later observed in high-frequency electron-spin resonance (ESR) and neutron scattering measurements~\cite{Asano2000,Kenzelmann2004,Zvyagin1998}.

The molecule-based chiral spin chain \cuchch\, (pym = pyrimidine) is structurally similar to the SG systems, but with an added twist: a 4$_1$ screw~\cite{Cordes2007}. A previous study employing X-ray diffraction, electron-spin resonance, magnetometry and heat-capacity measurements confirmed the chiral space group, a two-fold staggered component to the $g$ tensor and a field-induced gap~\cite{Liu2019}. However, it was found that in this case the gap is much smaller than that predicted by SG theory and has an unconventional linear-in-field dependence in contrast to the $H^{2/3}$ behavior expected for the SG model. Moreover, the observed magnetic excitations, while reminiscent of the breather modes seen in the non-chiral staggered chains, cannot be accounted for by the SG model~\cite{Liu2019}. 

A qualitative explanation for the departure of the \cuchch\ spin gap from existing theories was proposed based on differences between the chiral and non-chiral staggered structures~\cite{Liu2019}.  The screw symmetry introduces additional terms to the Hamiltonian, including a uniform DM interaction and a staggered field with a four-fold periodicity on each chain. These additional interactions have the potential to decouple the system from the two-fold staggered field and radically alter the development of the gap and the excitations. This could occur if, for example, in an applied field the spins adopt either an incommensurate helical magnetic order or a canted AFM state in which the moments lie in the plane defined by the applied field and the chain direction~\cite{Liu2019}. Later theoretical calculations predicted the development of spin-dimer order possibly coexisting with N\'{e}el order in this material~\cite{Furuya2020}. 

To examine these hypotheses, we present neutron diffraction, muon-spin relaxation and magnetometry data taken in applied magnetic field. We start by describing the crystal structure of the material and a minimal effective spin Hamiltonian. We then show that the neutron diffraction results taken at 50~mK indicate that long-range magnetic order emerges as a static field is applied and explain how this order develops as a result of the two-fold staggered field at each spin site. The muon and magnetometry results provide further information on the field at which long-range order sets in. We finish with a discussion of the implications of these results for the predictions mentioned above.

\section{Results and discussion}

\subsection{Crystal structure, Hamiltonian and zero-field behavior}

\begin{figure}[t]
 \includegraphics{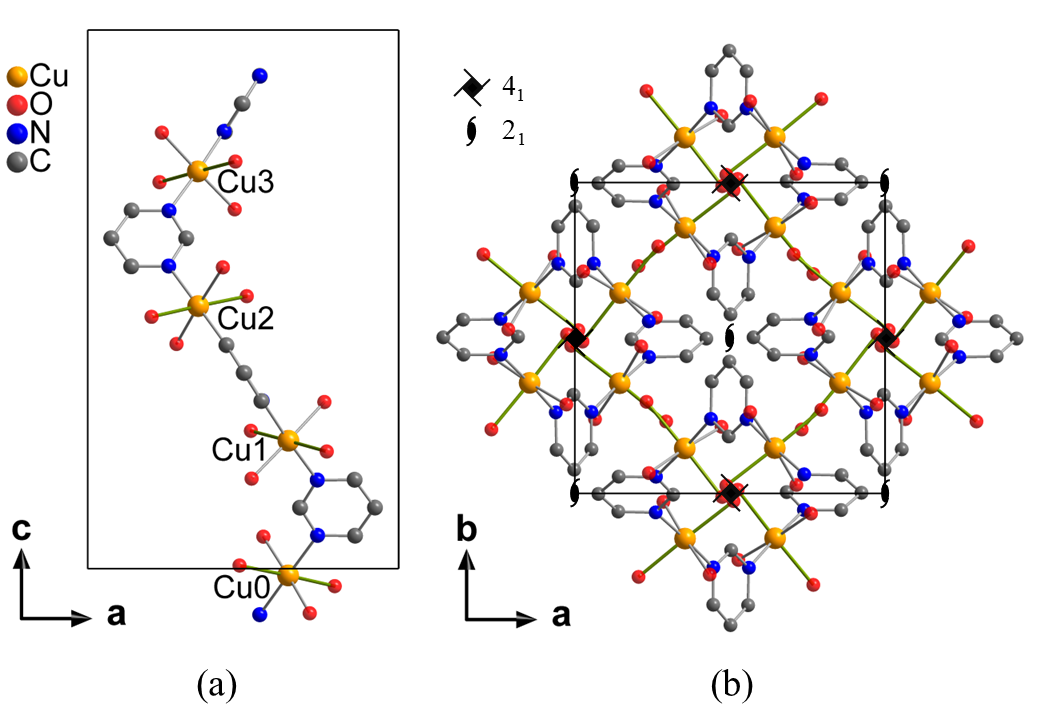}
 \caption{\label{fig:structure}(a) Crystal structure of the 4$_1$ Cu(pym)(H$_2$O)$_4$ chain and (b) four parallel chains related by the 2$_1$ screw axes in \cuchch. The Jahn-Teller elongated Cu--O bonds are shown as green sticks~\cite{Cordes2007,Liu2019} highlighting the four-fold-periodic staggering of the local spin environments. H atoms and non-coordinated SiF$_6$ and H$_2$O molecules are omitted for clarity.}
\end{figure}

As shown in Figure~\ref{fig:structure}(a), the Cu(II) ions in \cuchch\ form chains parallel to the $c$-axis via linking N---C---N moieties of pyrimidine, which mediate a strong AFM intrachain superexchange coupling, $J = 42$\,K~\cite{Liu2019}. The Cu(II) hexa-coordination is completed by four H$_2$O molecules, two of which occupy Jahn-Teller elongated axial positions. Along this direction, the local $g$ tensor assumes the largest value. In the $P4_12_12$ space group, the Cu(II) ions are related by the 4$_1$ symmetry element within the chain and by the 2$_1$ screw axis between chains, see Figure \ref{fig:structure}(b). As a consequence, the local environments of the nearest Cu(II) ions within a chain, and hence their local $g$ tensors, are related by 90$^\circ$ rotations about the crystallographic $c$ axis. The chains themselves are well separated from one another by H$_2$O molecules and SiF$^{-}_6$ counter-anions (not shown), leading to a minimum interchain Cu$\cdots$Cu distance of 7.4312(6)\,\AA\ at 150\,K. 

The minimal effective spin Hamiltonian appropriate for \cuchch\ was determined in Ref.~\cite{Liu2019}. Setting $\mathbf{Z}$ as the chain direction and applying a magnetic field $\mathbf{H}$, the Hamiltonian for a single chain can be written as
\begin{equation}
\label{hamiltonian}
\begin{aligned}
\hat{\cal{H}} & = \sum_i 
J\mathbf{\hat{S}}_i\cdot\mathbf{\hat{S}}_{i+1}
 + \mu_{\rm B}\mu_0\sum_i\mathbf{H}\cdot\mathbf{g_i}\cdot\mathbf{\hat{S}_i} \\
 & + \sum_i D_\mathrm{u}(\hat{S}_i^X\hat{S}_{i+1}^Y-\hat{S}_i^Y\hat{S}_{i+1}^X).\\ 
\end{aligned}
\end{equation}
where the first term describes the AFM exchange within the Cu–pym–Cu chain, $\mathbf{g_i}$ is the $g$ tensor at a particular spin site, and the last term arises from the DM interaction. 
The $g$ tensor can be split into three components: $g_{\rm u}$ corresponds to the uniform part of the $g$ tensor, while $g_{\rm 2s}$ is a small staggered component that repeats after every two nearest-neighbor Cu(II) ions. $g_{\rm 4s}$ is a much weaker staggered component with a four-fold period along the chain, i.e. the same periodicity as the unit cell. Using electron-spin resonance, the $g$ tensor has been experimentally determined to be (in the laboratory frame defined in Ref.~\cite{Liu2019}),
\begin{equation}
\label{Eqg}
\begin{aligned}
g_{i} &= g_{\rm u} + g_{\rm 2s} + g_{\rm 4s}\\
&= \begin{pmatrix}
2.21 & 0 & 0 \\
0 & 2.21 & 0 \\
0 & 0 & 2.10 
\end{pmatrix}
+ 0.12 \begin{pmatrix}
0 & (-1)^i & 0 \\
(-1)^i & 0 & 0 \\
0 & 0 & 0 
\end{pmatrix} \\
&+ 0.026 \begin{pmatrix}
0 & 0 & (-1)^i\delta_i \\
0 & 0 & \delta_i \\
(-1)^i\delta_i & \delta_i & 0 
\end{pmatrix}
\end{aligned}
\end{equation}
where $\delta_i = -1, +1, +1$ and $-1$ respectively for spins Cu0, Cu1, Cu2 and Cu3 along the chain (see Fig.~\ref{fig:structure} for label definitions). These components allow one to split Zeeman energy in the Hamiltonian into three terms: a uniform field, two-fold and four-fold staggered fields. It is the four-fold term that allows for the possibility of dimer order in this system~\cite{Furuya2020}. By symmetry, the DM interaction of a given chain can be decomposed into a uniform component parallel to the chain axis of magnitude $D_{\rm u}$, and a four-fold periodic staggered component perpendicular to the chain axis (not included in Eq.~\ref{hamiltonian}). 
It is important to note that space group $P4_12_12$ does not support a net DM interaction. The $2_1$ and 2 symmetry operators give the same configuration of staggered $D_{\rm s}$ on every chain, but opposite signs of the uniform $D_{\rm u}$ for the two chiral chains in each unit cell. 

In addition to these terms, it is possible that interchain coupling, characterized by an exchange energy $J_\perp$, plays a role in any magnetic order that develops in this system. Previously reported muon-spin rotation measurements have shown that at zero field \cuchch\, does not undergo a transition to long-range three-dimensional magnetic order, at least above 20\,mK~\cite{Liu2019}. Within the model of a conventional linear quantum chain, this would suggest that the exchange anisotropy is very small indeed; $J_\perp/J \lesssim 10^{-4}$\,K~\cite{Yasuda2005}. 
However the presence of a uniform DM interaction that alternates in sign on neighboring chains is known to compete with interchain exchange~\cite{Jin2017} and could suppress zero-field long-range order (LRO) in this system even in the presence of a higher $J_\perp$ value.

Clear evidence of zero-field order is also elusive in the non-chiral SG chain copper benzoate. 
By contrast, CDC does show magnetic order below 0.93\,K at zero field. 
Once a magnetic field is applied, one would expect LRO to develop alongside the energy gap in all these systems as the staggered fields should stabilize a preferred ground state for each chain. 
This is indeed observed in CDC, where field applied along one axis strengthens the exchange-induced LRO seen at zero field. However, for other directions, the staggered fields suppress the zero-field ordered state and cause a SG spin gap to open~\cite{Chen2007}.

\begin{figure*}
 \includegraphics[height=5cm]{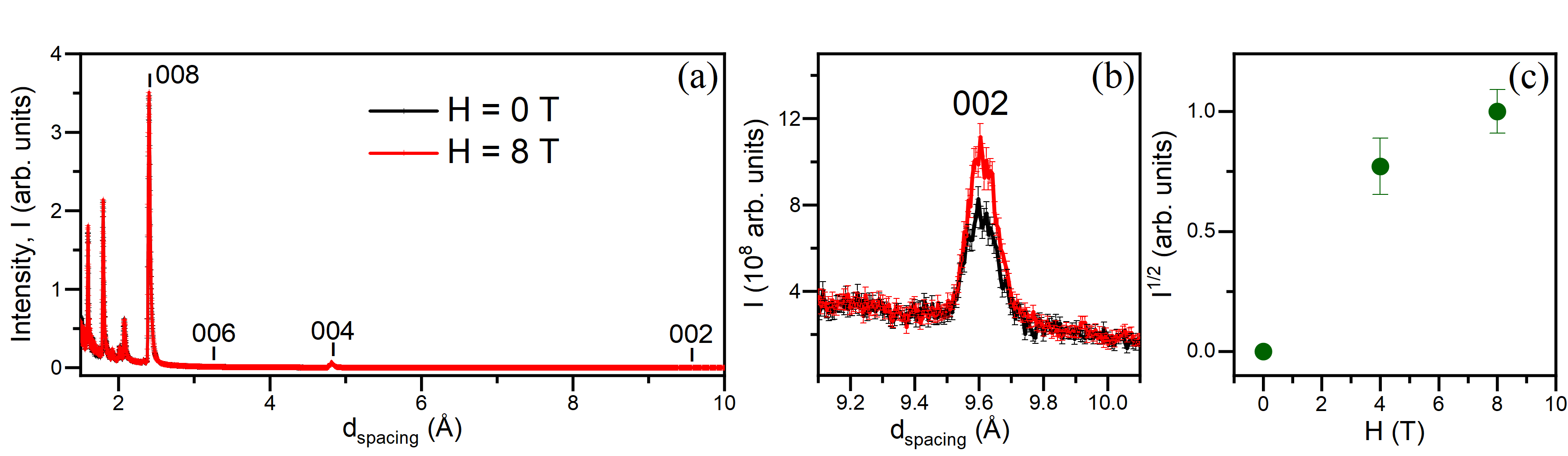}
 \caption{\label{fig:ndata} (a) Elastic neutron scattering diffractogram along the $(00l)$ direction for a single crystal of \cuchch\ at 50\,mK in zero magnetic field (black) and in a 8\,T magnetic field applied along the $[-1, 2, 0]$ direction (red). (b) Scattered intensity at $(002)$ showing that magnetic scattering occurs at 8\,T in addition to the multiple scattering observed at zero field. (c) Magnetic field dependence of the square-root of the magnetic scattering peak at $(002)$, which is proportional to the size of the ordered moment. These points are obtained by subtracting data collected at zero field from those obtained in applied field.} 
\end{figure*}

\subsection{Neutron diffraction}

In order to establish whether long-range magnetic order evolves in an applied field in \cuchch, we measured single crystal neutron diffraction on the WISH instrument at the ISIS Neutron and Muon Source, UK~\cite{Chapon2011}. The details of this and the other measurements described here are given in the Supplemental Material~\cite{SM}. This experiment poses some technical challenges. Magnetic scattering intensities are necessarily weak due to the low density of magnetic ions compared to typical inorganic materials. Moreover, Cu(II) moments are intrinsically small and strong quantum fluctuations typical of highly one-dimensional $S=1/2$ systems act to reduce the ordered moment even further~\cite{Shultz1996,Lancaster2006} (see discussion below).

Despite these difficulties, magnetic scattering in an applied field is clearly observed, demonstrating that the field induces long-range magnetic correlations in this material. Figure \ref{fig:ndata}(a) shows the measured diffraction intensity along the $(00l)$ direction at 50\,mK in zero magnetic field (black line) and in an 8\,T magnetic field applied along the $[-1,2,0]$ direction (red line). The field produces a measurable change in intensity along the $(00l)$ zone axis at a $d$-spacing of 9.6\,\AA, see Figure \ref{fig:ndata}(b). This position corresponds to the $(002)$ reflection, determined by indexing the diffraction pattern by the tetragonal cell with lattice parameters: $a=b=11.1513$\,\AA\ and $c=19.3206$\,\AA.  We note that a sizeable intensity is measured at this position even in zero magnetic field, despite the fact that long-range magnetic order is absent in zero field and nuclear intensities at $(00l)$ with $l = 4n + 2$ are extinct by space group symmetry. On rotating the sample over a small angular range this zero-field intensity was found to change dramatically and with an irregular dependence on angle. These characteristics are unique to multiple scattering, in which the Bragg condition is accidentally satisfied by two or more sets of crystal planes (we note that the sample orientation was selected to optimize measurement of the long d-spacing magnetic diffraction despite significant multiple scattering). Hence, the space group assignment holds, and the non-magnetic zero-field $(002)$ intensity can be subtracted from the high field data in the analysis that follows.

The magnetic intensity observed at the $(002)$ position is consistent a Gamma-point magnetic structure. Furthermore, the reflection condition $(00l)$ with $l = 4n + 2$ uniquely identifies spin components of opposite sign located on intra-chain Cu sites related by $4_1$-screw symmetry. The implied AFM exchange mediated by pyrimidine is consistent with the energy of the AFM interaction established by previous magnetometry and heat capacity measurements of this material~\cite{Liu2019}. The AFM component of the spins must coexist with a ferromagnetic (FM) component coupled to the magnetic field. Observing this FM component in our diffraction experiment is unlikely, given that it exactly coincides with nuclear diffraction intensity from the crystal structure. 

\begin{figure*}
 \includegraphics{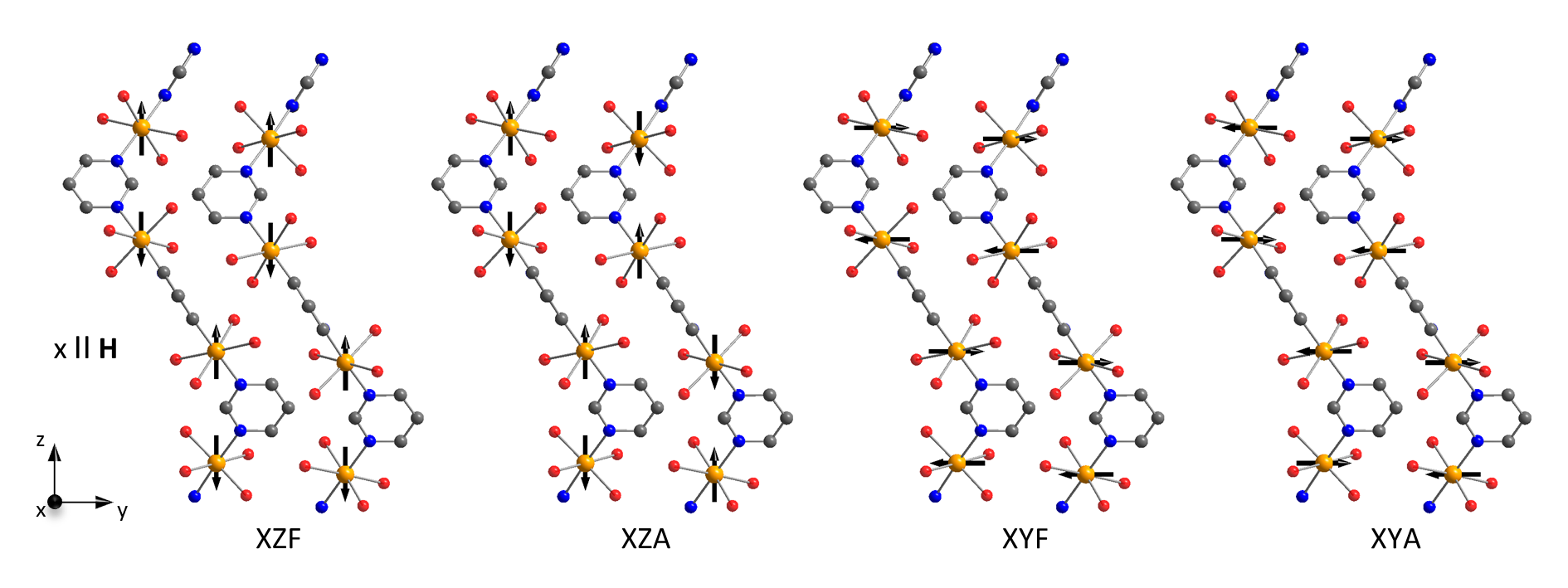}
 \caption{\label{fig:mstructures}Possible magnetic structures in applied magnetic field perpendicular to the chains of \cuchch .}
\end{figure*}

We define four possible antiferromagnetic structures, illustrated in Figure \ref{fig:mstructures}. Here, the small field-induced FM component necessarily lies parallel to the applied field (the X direction), while the intrinsic AFM component stabilized by exchange can lie either perpendicular or parallel to the chain axis along Y or Z, respectively. We label these two scenarios ‘XY’ and ‘XZ’. Secondly, the relative orientation of nearest-neighbor spins in neighboring chains may be FM or AFM, denoted as ‘F’ or ‘A’. Accordingly, we label these structures XYF, XYA, XZF and XZA. 

XZF and XZA can be immediately ruled out as candidates for the 8\,T magnetic structure, as antiferromagnetic moments aligned parallel to Z would give exactly zero neutron diffraction intensity at the $(002)$ reflection. The XYF and XYA magnetic structures can be distinguished by a $\cos^2(\epsilon l)$ and a $\sin^2(\epsilon l)$ dependence on the diffraction intensities, respectively, where $\epsilon$ is an irrational multiple of 2$\pi$ dependent upon the atomic fractional coordinates. However, separating these two requires a reliable measurement of the $(006)$ magnetic intensity for comparison, which is heavily suppressed by the magnetic form factor and is here lost to the background noise.

The size of the ordered magnetic moment $\mu_\textrm{Cu} = 0.174(15)\,\mu_{\rm B}$ obtained by fitting the XYF model against the intensity at (002), scaled using the nuclear Bragg intensities, is significantly reduced from the value expected for a Cu(II) ion in this environment. By contrast, the ordered moment found from fitting to the XYA magnetic model is large; 1.104(94)\,$\mu_\mathrm{B}$, consistent with an approximately full $S=1/2$ moment. A significant renormalization of the ordered magnetic moment is expected in highly isolated $S=1/2$ chains due to the effect of quantum fluctuations~\cite{Shultz1996} and it is reasonable to suppose the ordered moment remains reduced even at 8\,T, which is only 14\% of the saturation field in this material~\cite{Liu2019}. Hence, the derived moment size tends to support XYF as the best candidate for the 8\,T magnetic structure. 

Figure \ref{fig:ndata}(c) shows the square-root of the magnetic scattering at the $(002)$ peak, obtained by subtracting the data collected at zero field from that observed in applied field. Magnetic order is already present at 4\,T and it appears to strengthen as field increases to 8\,T.

The location of the ordered moments in the XY plane is contrary to the proposition put forward in Ref.~\cite{Liu2019} to account for the reduced size of the field-induced energy gap compared to the predictions of the SG model. It was suggested that in a magnetic field, the spins might adopt a canted structure in the XZ plane instead, which would effectively decouple the spins from the staggered field arising from the two-fold $g$-tensor component, which lies perpendicular to Z for the experimental field direction of $[-1, 2, 0]$. However, the results of our neutron diffraction experiment unambiguously show that the spins instead order within the XY plane defined by the applied and two-fold staggered fields, analogous to the situation expected for the non-chiral, staggered chains. This situation means that if the interchain interactions are small (as strongly suggested by the lack of evidence for long-range order in zero applied field) the two-fold staggered field should be responsible for the ordering between chains. 

\begin{figure}
\includegraphics[height=8.5cm]{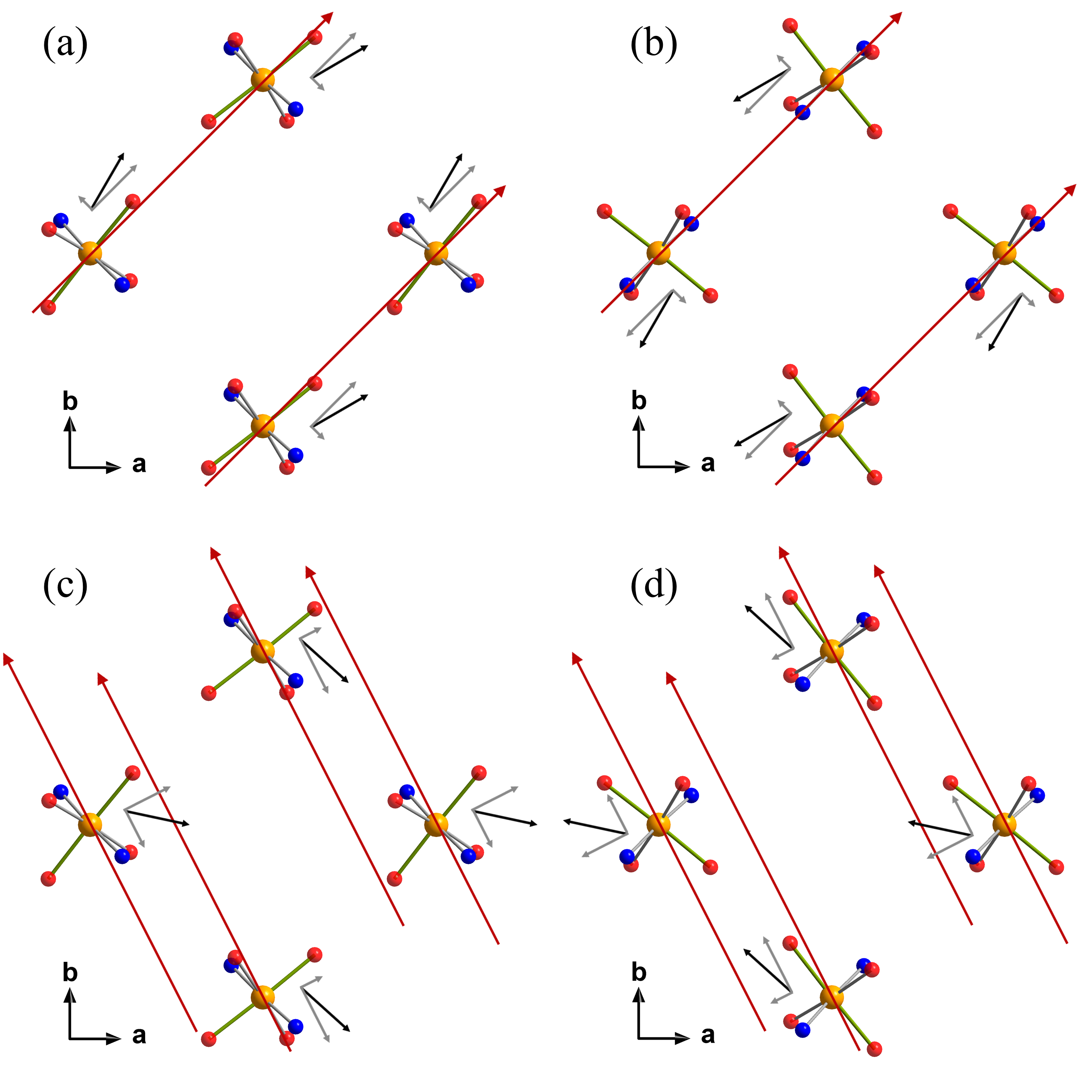}
 \caption{\label{fig:Hind} Cu(II) octahedral environments in four adjacent chains for Cu0 (a, c) and Cu1 (b, d) spins in the unit cell (see labels in  Figure~\ref{fig:structure}), as well as the two-fold staggered field (black arrows), decomposed into components parallel and perpendicular to an applied magnetic field (red arrows). (a, b) applied field parallel to [1 1 0]. (c, d) applied field parallel to $[-1, 2, 0]$, as used for the neutron diffraction experiments.}
\end{figure}

To understand how magnetic order arises, it is helpful to shift away from the laboratory XYZ frame of Ref.~\cite{Liu2019} to the crystallographic $abc$ frame. The orientation of the local Jahn-Teller (JT) axes (and hence the local $g$ tensor) rotates by $90^\circ$ about the $c$-axis on moving from one Cu(II) site to the next along the chiral chains. This is illustrated in Fig. 4 for Cu(II) ions in four adjacent chains. The left-hand panes (a, c) show the situation for Cu0 spins along the chain, while the right-hand panes (b, d) show the situation for Cu1 (see Figure~\ref{fig:structure} for label definitions). The JT axes of the Cu0 ions lie approximately parallel to the $[1,1,0]$ direction, with a small offset angle of $\alpha = \pm7.5^\circ$ when projected into the $ab$-plane. The sign of $\alpha$ alternates between adjacent chains, as shown. For Cu1, the JT axes lies approximately parallel to the $[\bar{1},1,0]$ direction, again with a $\pm7.5^\circ$ offset that alternates between chains. Cu2 and Cu3 ions have the same JT-axis orientations as Cu0 and Cu1, respectively, when projected into the $ab$-plane, which is a sufficient consideration for the discussion that follows.

It is the orientation of the local $g$ tensor that dictates the effective field at a particular site. For example, if the JT axis is aligned exactly parallel to the applied field, the effective field at that site will be boosted, but if the JT axis is exactly perpendicular to the applied field, the effective field will be reduced. The alternation of JT directions described above therefore gives rise to the two-fold staggered field ($h_{\rm 2s}$), superposed on an average uniform field, that alternates in direction from site to site along the chains. Using the experimentally derived $g$ tensor shown in Equation~\ref{Eqg}, it can be shown that the two-fold staggered field is given by,
\begin{equation}
    h_{\rm 2s} = \pm 0.12 
    \begin{pmatrix}
H_{a}\sin(2\alpha) + H_{b}\cos(2\alpha)\\
H_{a}\cos(2\alpha) + H_{b}\sin(2\alpha)\\
0
\end{pmatrix}
\end{equation}
where $h_{\rm 2s}$ is positive for Cu0, alternating in sign from site to site along the chain, $[H_{a}, H_{b}, 0]$ is the direction of the applied field in the $ab$ plane. For $\alpha \neq 0$, an applied field can never be aligned exactly parallel or perpendicular to all JT axes at once. Hence, $h_{\rm 2s}$ will always develop components parallel and perpendicular to the applied field. The antiferromagnetic moments (or staggered magnetization) will orient perpendicular to an applied field to maximize the magnetic susceptibility, however this is not enough to define a relationship between the chains. In fact, the relative orientation of moments in adjacent chains will be determined by these perpendicular components of $h_{\rm 2s}$, which are hence responsible for stabilizing long-range order.

Figure 4(a) and (b) illustrate this scenario. Here the applied field (red arrows) is applied parallel to $[1,1,0]$. The two-fold staggered fields are drawn as black arrows next to their respective Cu(II) octahedra, as calculated using Equation 3, and decomposed into components parallel and perpendicular to the applied field (gray arrows). In the absence of the $\alpha$ offset angle, $h_{\rm 2s}$ would be exactly parallel (at Cu0) and antiparallel (at Cu1) to the applied field. However, the $\alpha = \pm7.5^\circ$ offset introduces an additional component of $h_{\rm 2s}$ perpendicular to the applied field, which for this field direction alternates in direction between adjacent chains and hence promotes an antiferromagnetic order corresponding to the XYA magnetic structure in Fig.~\ref{fig:mstructures}. (Note that $h_{\rm 2s}$ is inclined at an angle of $2\alpha$ with respect to the applied field, see Equation 3.) Figure 4(c) and (d) shows the situation for field applied parallel to $[-1,2,0]$, which is the direction used in the neutron experiments. In this case, the perpendicular components of $h_{\rm 2s}$ no longer alternate between adjacent chains and so act to promote the XYF ferromagnetic ordering, as was deduced from the neutron results. In fact, it can be shown that the XYF arrangement will be favored provided that $\cos(4\alpha) + \cos(4\phi) > 0$, where $\phi$ is the azimuthal angle that defines the direction of the applied field in the XY plane. In other words, the XYF structure will always occur unless the field is applied within $\pm\alpha$ of the $\langle 110\rangle$ directions. 

Finally, we note that for fields applied parallel to the chain (crystallographic $c$-axis), $h_{\rm 2s}$ is exactly zero and long-range order will not be stabilized by this mechanism. 

Summarizing the neutron diffraction results, commensurate antiferromagnetic order is observed in fields of 4 and 8\,T, with the spins aligning in the plane defined by the applied and two-fold staggered fields. For the experimental field direction, nearest-neighbor spins on adjacent chains are ferromagnetically aligned.    

\subsection{Muon-spin rotation and magnetometry}


To understand at what point field-induced magnetic order begins, transverse-field (TF) muon-spin rotation ($\mu^+$SR) measurements were performed as described in the Supplemental Material~\cite{SM}.

 In the TF $\mu^+$SR
geometry \cite{blundell2022} the externally applied field $B_{0}$ is
directed perpendicular to the initial muon spin direction.
Muons
precess about the total magnetic field $B$ at the muon site, which
includes the contribution of the internal magnetic field from local spins. The
observed property of the experiment is the time evolution of the muon-spin polarization $P_x(t)$, which is related to the
distribution $p(B)$ of local magnetic fields at the muon sites across the sample volume
by means of a Fourier transform,
\begin{equation}
P_{x}(t) = \int\mathrm{d}B\,p(B)\cos(\gamma_{\mu}Bt),
\label{eq:muon1}
\end{equation}
where $\gamma_{\mu}$ is the muon gyromagnetic ratio.
In the high magnetic field limit, the muon-spin relaxation rate
(and hence the width of the features seen in the Fourier
transforms of the spectra) are determined by the magnetic field correlations along the direction of the applied
magnetic field. The signal we observe is a consequence of contributions from the
local fields in the sample, which will include the dipole field due to
the local moments along with the hyperfine field due to electron
density at the muon sites \cite{blundell2022}.
In a magnetically ordered system muons will stop both in the bulk ordered regions
and in domain walls and at surfaces, where the internal field is nonuniform.
There will also be a contribution from muons that stop in the silver sample
holder and possibly other regions outside the sample. 

\begin{figure}
\includegraphics[width=\columnwidth]{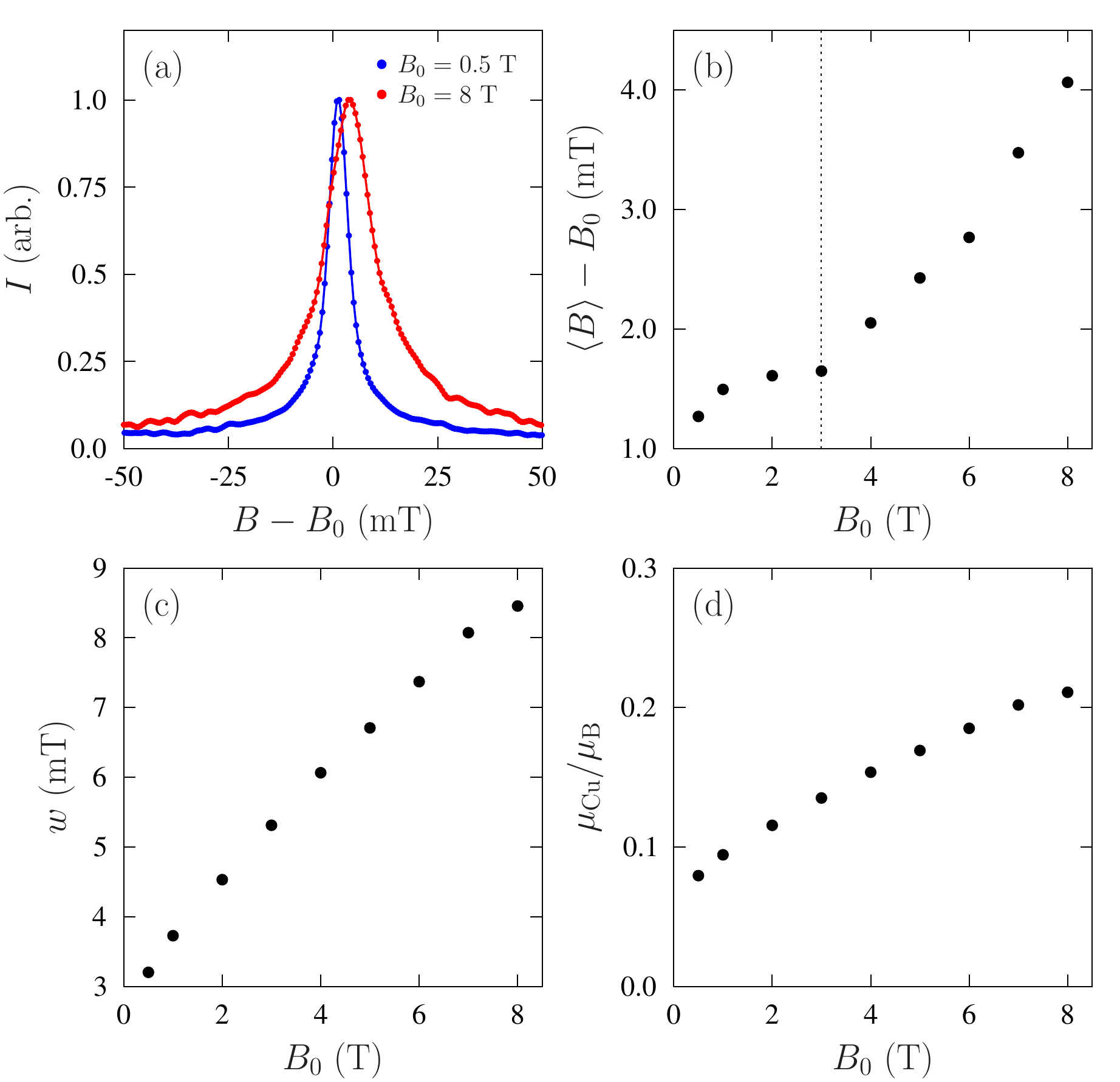}
\caption{(a) Example 0.1~K TF $\mu^+$SR Fourier transform spectra measured at
  0.5 and 8~T.
  Fits of the spectra to Eq.~\ref{eq:muon2}
  showing (b) the average shift $\langle B\rangle-B_{0}$ as a
  function of applied field, and (c) the broadening $w$. The dotted line shows the proposed
  transition region. (d) Ordered moment size derived from
  muon site modelling for the XYF magnetic structure.\label{fig:muonfig2}}
\end{figure}

TF $\mu^+$SR spectra were measured on a polycrystalline sample in applied
 fields 0.5~T~$\le~B_0 \le$ 8.0~T at a fixed temperature of $T = 0.1$~K.
Example Fourier-transform spectra are shown in
Fig.~\ref{fig:muonfig2}(a), in which we see
a
single broad peak at all applied fields.
As the externally applied field increases the peaks become broader and shift
to higher average magnetic field.
We quantify these features by fitting each peak to a Lorentzian of the form
\begin{equation}
I(B) = \frac{A w^2}{\left[{\left(B - \langle B \rangle\right)}^2 + w^2\right]} + A_0,
\label{eq:muon2}
\end{equation}
where $B$ is the measured field, $A$ is the height of the peak, $w$ its half-width at half-maximum,
and ${A_0}$ is a baseline contribution.
As shown in Fig~\ref{fig:muonfig2}(c),
$w$ is seen to increase smoothly as a function
of applied field.
In contrast, the shift of the average from the applied field $\Delta B = \langle B\rangle - B_{0}$
[Fig.~\ref{fig:muonfig2}(b)] stays approximately
constant up to $\approx 3$~T, when it starts to increase roughly
linearly.

The marked
discontinuity in the gradient of $\Delta B$ is suggestive of a field-induced
transition in magnetic  behavior that takes place around
$B_{\mathrm{c}}=3$~T. This would most likely reflect either a phase
transition from a disordered configuration for $B_{0}<3$~T to long-range order, or a transition
between two different magnetically-ordered phases.
In either case, we might  also expect to see a discontinuity in
the width of the measured distribution $w$.
However, the smooth increase in $w$ that we observe indicates that the distribution of
magnetic fields at the muon sites does not change discontinuously
across the measured field range.
This might be  consistent with a reordering between two subtly
different magnetic structures that do not involve a significant
change in ordered moment size. 
It can also be reconciled with
the scenario involving the onset of LRO at $B_{0}=3$~T, which would
imply that the correlations either side of the phase boundary are
similar, with short-range ordered regions occurring for $B_{0}<3$~T that
lock into LRO at high fields without a significant change in the local
distribution of fields.  

In order to check whether the muon results are consistent with the
magnetically ordered
structures proposed on the basis of neutron diffraction, we modelled  the spectra, assuming only a
dipolar contribution to the local magnetic field. The simulation involved computing
the candidate muon sites in the system (using density functional
theory techniques \cite{blundell2022}) along with modeling the dipole field distribution
resulting from the candidate ordered magnetic structures.
A large number of candidate muon sites were identified, with only the
lowest-energy species retained.
We then computed the magnetic field distributions expected from these
muon sites in a polycrystalline sample, but allowed the magnitude of
the ordered moments to serve as a fitting parameter. 
Using this approach it was not possible to recreate the observed magnitude of
the shifts $\Delta B$, which resulted in unrealistically large values
of the ordered moments for all candidate magnetic structures. This
suggests that the shifts do not arise simply from the dipole field, and
instead reflect one of the contributions not taken into account, such
as a hyperfine shift. 
In contrast, the distribution width $w$ can be successfully modeled
for both the
proposed  XYF and XYA magnetic structures, 
with physically-realistic values of the ordered moment providing good
agreement between the data and simulation. 
For both magnetic structures, the magnitude of the ordered moment
shows
a tendency to increase with increasing applied field.
The extracted moments lie in the
range $0.1\lesssim \mu/\mu_{\mathrm{B}}\lesssim 0.2$, as shown in the
Fig.~\ref{fig:muonfig2}(d), where we have computed the ordered moment
that would be consistent with the data, assuming an XYF structure is realized at all
values of applied magnetic field. 

The measured behavior can also be compared with that observed in other coordination polymer materials measured using the same technique. In Cu(pyz)(gly)(ClO$_{4}$), for example \cite{lancaster2018},  the observed shifts $\Delta B$ correlate with the known transitions at phase boundaries. In that case, the boundary between a phase of interacting singlets and the XY LRO phase shows a discontinuity in the magnitude of $\Delta B$, as might be expected, since this involved the discontinuous appearance of magnetic moments. By contrast, the transition between a  field-polarized state and an XY-ordered state involves a change in
gradient of the shift, much as we see here. This further implies that our results, involving a change in gradient of $\Delta B$,
are consistent with a transition between a magnetically ordered phase and one with correlations imposed by the polarization that results from an applied field. 

In summary,
the behavior of the measured lineshift $\Delta B$, which changes from the field-independent at low
 $B_{0}$ to linearly increasing for $B_{0}>3$~T, is suggestive
 of a qualitative change in the underling magnetic state, such as that
 from a magnetically disordered phase for
 $B_{0}<3$~T to field-induced order above $B_{\mathrm{c}}$.
 The ordered moment
 derived from fits to the linewidth are consistent with both the proposed
 XYF or the XYA magnetically-ordered structures across the field range, with an ordered moment
 of $\approx 0.2~\mu_{\mathrm{B}}$ at 8~T for the XYF structure, in agreement with the neutron diffraction result.

\begin{figure}
 \includegraphics[width=\columnwidth]{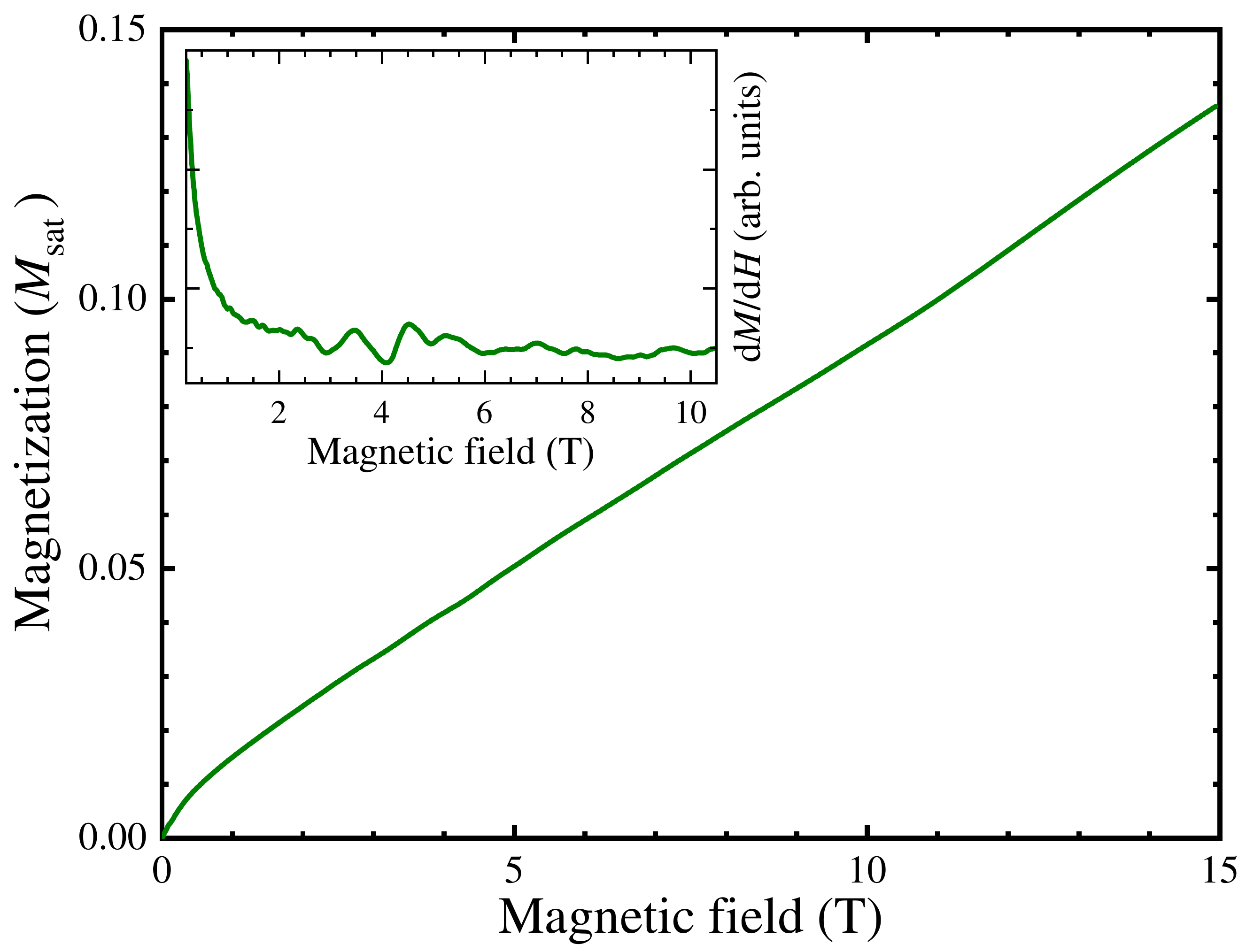}
\caption{Magnetization of \cuchch\ measured in pulsed magnetic fields at $T = 0.6$~K for fields applied perpendicular to the chain. The data are shown as a fraction of $M_{\rm sat}$, the saturation magnetization~\cite{Liu2019}. Inset: differential susceptibility in the low-field region.\label{fig:mag}}
\end{figure}

Fig.~\ref{fig:mag} shows the magnetization $M$ of \cuchch\ measured at pumped $^3$He temperatures in pulsed magnetic fields. The data were obtained by coaligning two single crystals with the field applied perpendicular to the chain axis. To eliminate any systematic errors arising due to mechanical vibrations, a total of 35 field pulses were performed in two separate magnets with different peak fields between 10 and 20~T. The figure shows the average of these datasets. Broadly, the $M(H)$ data appear to rise quickly from zero field with a decreasing gradient up to 2--3~T. At higher fields up to 15~T the data continue to increase but with a roughly constant gradient. In addition, a subtle modulation is just discernible in the 3--5~T region. This modulation becomes clearer on differentiation of the $M(H)$ data (shown in inset) and appears as a double-kink feature near the point where gradient levels off. Similar features have recently been reported in the same field region in measurements of the electric polarization of this material~\cite{Blockmon2024}.  
The change in behavior of $M(H)$ is close to the position at which the $\mu^+$SR data show a shift in the internal field, adding further support for a field-induced transition from short-range correlations to long-range order. 

\section{Conclusions}

In summary, our experimental data demonstrate the development of long-range magnetic order in \cuchch\ under applied magnetic field. The transverse-field muon measurements described above are indicative of a phase transition to a magnetically ordered state at 100~mK in an applied field of around 3~T. This is in agreement with the neutron diffraction results which show the presence of long-range commensurate antiferromagnetic order at both 4 and 8~T. 

Using density-matrix renormalization group and non-Abelian bosonization, Ref.~\cite{Furuya2020} focused on the four-fold staggered field term in the Hamiltonian and predicted a coexistence of dimer and Néel order developing in applied fields. However, this is not what is indicated by the neutron data. Instead, XYF order is found, in which the spins lie in the plane defined by the applied field and the two-fold staggered field, precluding dimer formation. In addition, the observed ferromagnetic alignment of neighboring chains is consistent with the direction of the two-fold staggered field. This is exactly the situation expected for the non-chiral, alternating staggered chains, where the classical picture of field-induced order is a canted phase caused by the combination of the interchain exchange and the applied and two-fold staggered fields. It is the combination of these three energy considerations that map, via bosonization, on to the SG model. Our results therefore suggest that the SG equation should remain a good approximation for the low energy properties in \cuchch\ (at least for fields applied perpendicular to the chain). In this case, the departure of both the size and field dependence of the field-induced energy gap from the predictions of the SG model remains a perplexing issue that needs to be explained.

Having ruled out effects of the four-fold staggered field, we suggest that the departure from SG physics may be due to the uniform component of the DM interaction of a given chain, which alternates between nearest-neighbour chains as required by the symmetry of the crystal structure. This interaction also provides a natural explanation for the remarkable absence of long-range magnetic order at 20 mK. The uniform DM terms promote incommensurate helical antiferromagnetic order with opposing chirality on neighboring chains, which competes with the interchain exchange interactions that favor commensurate collinear order at finite temperature. The transition at 100 mK and 3 T may then occur when the energy of the two-fold staggered field critically exceeds that of the DM interaction. It is clear that further experimental and theoretical work is needed to fully understand the properties of this chiral quantum chain.

\section*{Acknowledgments}
We thank J. Liu and J. Villa for useful discussions, as well as T. Orton and P. Ruddy for technical assistance. This project has received funding from the European Research Council (ERC) under the European Union’s Horizon 2020 research and innovation programme (Grant Agreement No. 681260) and EPSRC (Grant No. EP/N024028/1). A portion of this work was performed at the National High Magnetic Field Laboratory (NHMFL), which is supported by National Science Foundation Cooperative Agreement Nos. DMR-1644779 and DMR-2128556 and the Department of Energy (DOE). J.S. acknowledges support from the DOE BES program “Science at 100 T,” which permitted the design and construction of much of the specialized equipment used in the high-field studies. The work at EWU was supported by the NSF through grant no. DMR-2104167. Part of this work was carried out at the Swiss Muon Source, Paul Scherrer Institute, and we are grateful for the provision of beam time and to R. Scheuermann for experimental support. AHM acknowledges support from STFC-ISIS and EPSRC. ZG acknowledges support from the Swiss National Science Foundation (SNSF) through SNSF Starting Grant (No. TMSGI2 211750). Work at the University of Oxford by BMH and SJB was funded by UK Research and Innovation (UKRI) under the UK government's Horizon Europe guarantee funding (Grant No. EP/X025861/1). Data presented in this paper will be made available at XXX. The neutron diffraction data are available at https://doi.org/10.5286/ISIS.E.RB1910030. For the purpose of open access, the author has applied a Creative Commons Attribution (CC-BY) license to any Author Accepted Manuscript version arising from this submission.


\bibliography{bibliography.bib}


\end{document}